\documentclass[12pt,preprint]{aastex}
\usepackage{color}
\usepackage{graphics,graphicx}
\usepackage{rotating}

\newcommand{\lapprox} {\, \lower3pt\hbox{$\sim$}\llap{\raise2pt\hbox{$<$}}\,}
\newcommand{\gapprox} {\, \lower3pt\hbox{$\sim$}\llap{\raise2pt\hbox{$>$}}\,}

\begin{document}

\title{Turbulent pitch-angle scattering and diffusive transport of hard-X-ray producing electrons in flaring coronal loops}

\author{Eduard P. Kontar\altaffilmark{1}, Nicolas H. Bian\altaffilmark{1},  A. Gordon Emslie\altaffilmark{2}, \& Nicole Vilmer \altaffilmark{3}}

\altaffiltext{1}{School of Physics \& Astronomy, The
University of Glasgow, G12 8QQ, Scotland, UK \\
eduard@astro.gla.ac.uk}

\altaffiltext{2}{Department of Physics \& Astronomy,
Western Kentucky University, Bowling Green, KY 42101 \\
emslieg@wku.edu}

\altaffiltext{3}{LESIA, Paris Observatory, Meudon, France}

\begin{abstract}
Recent observations from {\em RHESSI} have revealed that the number of non-thermal electrons in the coronal part of a flaring loop can exceed the number of electrons required to explain the hard X-ray-emitting footpoints of the same flaring loop. Such sources cannot, therefore, be interpreted on the basis of the standard collisional transport model, in which electrons stream along the loop while losing their energy through collisions with the ambient plasma; additional physical processes, to either trap or scatter the energetic electrons, are required. Motivated by this and other observations that suggest that high energy electrons are confined to the coronal region of the source, we consider turbulent pitch angle scattering of fast electrons off low frequency magnetic fluctuations as a confinement mechanism, modeled as a spatial diffusion parallel to the mean magnetic field. In general, turbulent scattering leads to a reduction of the collisional stopping distance of non-thermal electrons along the loop and hence to an enhancement of the coronal HXR source relative to the footpoints. The variation of source size $L$ with electron energy $E$ becomes weaker than the quadratic behavior pertinent to collisional transport, with the slope of $L(E)$ depending directly on the mean free path $\lambda$ again pitch angle scattering. Comparing the predictions of the model with observations, we find that $\lambda \sim$$(10^8-10^9)$~cm for $\sim30$~keV, less than the length of a typical
flaring loop and smaller than, or comparable to, the size of the electron acceleration region.
\end{abstract}

\section{Introduction}\label{intro}

One of the central ingredients of a solar flare is the efficient acceleration of electrons to suprathermal energies.  These electrons can be observed {\it in situ}, when they escape the Sun into interplanetary space \citep[see, e.g.,][]{1985SoPh..100..537L,2007ApJ...663L.109K}, or remotely, through the gamma-rays, X-rays and radio waves they emit \citep[see, e.g.,][for recent reviews]{2011SSRv..159....3D,2011SSRv..159..107H,2011SSRv..159..301K,2011SSRv..159..167V,2011SSRv..159..357Z}. In the commonly-adopted ``footpoint'' flare scenario
\citep[e.g.,][]{1958PhRvL...1..205P, 1968IAUS...35..471S,1968ApJ...151..711A,1969ARA&A...7..149S,1971SoPh...18..489B,1972SvA....16..273S}, electrons accelerated in the corona spiral along guiding magnetic field lines, losing a relatively insignificant amount of energy in the somewhat tenuous coronal environment.  They then reach the high plasma density regions of the lower solar atmosphere, where they emit the bulk of their X-rays via electron-ion bremsstrahlung and also lose the bulk of their energy through electron-electron Coulomb collisions.

Recent observations from {\em RHESSI} \citep{2002SoPh..210....3L} have provided unprecedented hard X-ray imaging spectroscopy data, allowing the study of the spatial structure of hard X-ray-emitting regions in solar flares. Such observations usually reveal the presence of coronal sources at energies $\lapprox 20$~keV and footpoint sources at higher energies $ \gapprox 30 $~keV \citep[e.g.,][]{2002SoPh..210..229K,2003ApJ...595L.107E,
2006A&A...456..751B,2007ApJ...665..846P,2011ApJ...740...46H}. Observations from {\em RHESSI}, both alone \citep[e.g.,][]{2002SoPh..210..383A,2010ApJ...717..250K} and more recently in combination with {\em SDO} data \citep{2012ApJ...760..142B},
support the ``footpoint'' scenario outlined above, indicating not only that photons of higher energy are emitted lower in the solar chromosphere but also suggesting a decrease in the size of HXR sources with depth that is consistent with the convergence of the guiding magnetic field lines as they penetrate
into the chromosphere.  Further, measurements of the difference in hard X-ray spectral index between footpoint and coronal sources suggest that the electron distribution spectrum in the corona is softer than that in the footpoints \citep[e.g.][]{2003ApJ...595L.107E,2006A&A...456..751B}.  Since the collisional energy loss rate is a decreasing function of energy, collisions lead to a hardening of the local electron spectrum.
Thus the relative hardness of footpoint sources relative to coronal sources in the same event lends additional support
to models that invoke collisional losses of the electrons in the loop plasma.

The most intense X-ray sources are associated with a high plasma density and hence a high collisional loss rate.  Indeed, for a sufficiently high ambient density, a coronal X-ray source region can be considered as a ``thick target,'' with the accelerated electrons remaining mostly confined within the coronal region. \citet{2008ApJ...673..576X}, \citet{2011ApJ...730L..22K}, and \citet{2012A&A...543A..53G} have shown that the extent of such sources parallel to the axis of the coronal loop grows with photon (or electron) energy. Since the collisional stopping distance of an electron of energy $E$ scales as $E^2$, such observations are broadly consistent with a model in which accelerated electrons stream along a loop of roughly uniform density without being significantly scattered.

More generally, the extent of a hard X-ray source is controlled by the confinement properties of non-thermal particles within the magnetized plasma in which they propagate.  Recent {\em RHESSI} analysis of HXR-producing electrons \citep{2013A&A...551A.135S} shows that the number of electrons above 30~keV in the coronal source is larger than that in the footpoints by a significant factor (between $\sim$2 and $\sim$8), suggesting a mechanism for enhanced entrapment of electrons in the loop top, possibly through either magnetic mirroring or turbulent pitch-angle scattering.

Efficient pitch angle scattering is a common requirement for stochastic acceleration during flares
\citep[e.g.,][for recent reviews]{2012SSRv..173..535P,2012ApJ...754..103B}. Moreover, the presence of magnetic fluctuations in flaring loops is suggested by the increase of loop width with energy revealed by {\em RHESSI} observations \citep{2011ApJ...730L..22K, 2011A&A...535A..18B}. The effects of turbulent pitch angle scattering, which may lead to diffusive transport in the limit of strong scattering, have been considered in the solar flare literature \citep[e.g.][]{1982ApJ...257..354H,1991ApJ...374..369B,2002SoPh..211..135S,2007A&A...465..613S} and used in the interpretation of solar flare
observations \citep[e.g.][]{1998A&A...334.1112J,2013ApJ...768..190F}, but no quantitative conclusions about the strength of
pitch-angle scattering with direct observational comparisons with HXR observations have been made.
The diffusion approximation for particle transport has also been used by many authors to explain the confinement of cosmic
rays and interpret synchrotron sources in the Galaxy \citep[e.g.,][]{1963ICRC....3..301G,1968PhRvL..20..752J}.

In this paper, we consider the influence of magnetic fluctuations on parallel electron transport in a flaring loop and we infer how HXR source sizes and spectra are affected by pitch angle scattering. Further, we derive an analytic expression for the energy-dependent source extent in the limit of strong pitch angle scattering when the parallel transport becomes diffusive.  We compare this expression with both the predictions of a purely collisional transport model.  We find that the {\em RHESSI} HXR observations are consistent with relatively weak parallel scattering, with an electron-scattering mean free path in the range $(10^8-10^9)$~cm.  Models that invoke mean free paths smaller than $\sim$$10^8$~cm (or equivalently the electron
isotropization times shorter then $10^{-2}$~s) are difficult to reconcile with the data.

\section{Diffusive parallel transport of energetic electrons}

The spatio-temporal evolution of the electron distribution function parallel to the background magnetic field $\mathbf{B}_{0}$ in a medium of density $n$ (cm$^{-3}$) is described by a one-dimensional Fokker-Planck equation
\begin{equation}\label{Eq:fp}
 \frac{\partial f}{\partial t}+\mu \, v \,  \frac{\partial f}{\partial z}
 =\frac{2K n(z)}{m_e^2} \, \frac{\partial }{\partial v}\left(\frac{f}{v^2}\right)+
\frac{\partial }{\partial \mu}\left(D_{\mu\mu} \, \frac{\partial f}{\partial \mu}\right) +S(v,\mu,x,t)\,\,\, ,
\end{equation}
where $f(z,\mu,v,t)$ is the electron distribution function (electrons~cm$^{-1}$~[cm~s$^{-1}$]$^{-1}$),
normalized to the electron number density: $\int \!\! \int f \, d\mu \, dv = n$, $v$ (cm~s$^{-1}$) is the speed
of the particle, $\mu$ is the cosine of the particle pitch angle relative to the guiding magnetic field ($z$-direction)
and $z$~(cm) is the distance from the top of the loop. The collisional parameter $K=2\pi e^4\Lambda$,
where $e$ is the electronic charge (e.s.u) and $\Lambda$ the Coulomb logarithm,
and $m_e$ (g) is the electron mass.
The equation (\ref{Eq:fp}) describes one dimensional propagation of non-thermal electrons along magnetic
field lines. The first term on the right hand-side describes energy losses due to binary collisions, while
the second term on the rhs of Equation (\ref{Eq:fp}) pitch angle scattering of electrons. The last term
$S(v,\mu,x,t)$ is the source term of electrons describing injection/acceleartion of particles.

The pitch-angle diffusion coefficient

\begin{equation}
D_{\mu\mu}=D_{\mu\mu}^{(C)}+D_{\mu\mu}^{(T)} \,\,\, ,
\end{equation}
in general consists of a collisional and a turbulent part. The collisional term is given by \citep[e.g.,][]{1983hppv.book.....G,1986CoPhR...4..183K}

\begin{equation}
D_{\mu\mu}^{(C)}=\frac{(1+{\overline{Z^2}}) K n(z)}{m_e^2} \, \frac{1}{v^3} \, (1-\mu^2) \,\,\,  ,
\label{eq:D_CC}
\end{equation}
where the factor $(1+{\overline{Z^2}})$ takes into account both electron-electron scattering and scattering on ions, with mean square atomic number ${\overline{Z^2}}$.
The presence of magnetic fluctuations inside the loop leads to an additional turbulent
contribution $D_{\mu\mu}^{(T)}$ (see Appendix as an example).
The {\it mean free path} $\lambda$ of a particle undergoing pitch angle
scattering is \citep[e.g.,][]{1989ApJ...336..243S}

\begin{equation}\label{eq:lambda_QLT}
\lambda \equiv\frac{3v}{8} \int _{-1}^{1} \frac{(1-\mu^2)^2}{D_{\mu\mu}^{(T)}} \, d\mu \,\,\, .
\end{equation}
In general, the mean free-path $\lambda$ could have a complicated dependency on speed $v$ depending on the spectral energy density
of the turbulence.  Since, for flaring plasma $D_{\mu\mu}^{(T)}$ is essentially unknown, we shall assume for simplicity that the mean free path $\lambda$ given by Equation (\ref{eq:lambda_QLT}) is a constant. Using this parameter $\lambda$ as the measure of pitch angle scattering,
we can quantify the characteristic pitch angle scattering timescale $\sim \lambda/v$ of electrons with speed $v$
and hence determine the importance of pitch angle scattering in flaring loops.

When pitch-angle scattering is strong enough, in the sense that $D_{\mu\mu}^{(T)} \, t \gg 1$, then pitch angle diffusion leads

to a flattening of the distribution function in $\mu$ over time $t$, i.e., an isotropization of the electron distribution,
so that $\partial f/\partial \mu\rightarrow 0$. In this limit, the operator describing ballistic transport
becomes (on average) a spatial diffusion parallel to the guiding field:

\begin{equation}
\mu v \, \frac{\partial f}{\partial z}\rightarrow D_{zz} \, \frac{\partial ^{2} f}{\partial z^{2}} \,\,\, ,
\end{equation}
and other processes (such as collisional losses) proceed at the same (energy-dependent) rate as they would in the absence
of scattering. The spatial diffusion involves an average of the pitch angle diffusion over pitch angles, according to

\begin{equation}
D_{zz} =\frac{v^2}{8}\int \limits _{-1}^{1} \frac{(1-\mu ^2)^2}{D_{\mu\mu}^{(T)}} \, d\mu = \frac {\lambda v}{3} \,\,\, .
\label{eq:D_zz1}
\end{equation}
and the collisional transport process can be modeled by

\begin{equation}\label{Eq:fp1}
 \frac{\partial f}{\partial t}
 =\frac{2K n(z)}{m_e^2} \, \frac{\partial }{\partial v} \, \left(\frac{f}{v^2}\right)+
D_{zz} \, \frac{\partial ^{2} f}{\partial z^{2}} \,\,\, .
\end{equation}

Although collisional pitch angle scattering does produce spatial diffusion of {\it thermal} electrons and can contribute to scattering of electrons,
it rather weakly affects the spatial transport of non-thermal electrons. The main reason is that the collisional pitch-angle scattering time is approximately the same as the energy loss time, as evident from Equation (\ref{Eq:fp}).  A crucial point, therefore, is that for pitch-angle scattering to be significant, it must operate on a time scale much less than the Coulomb collision time $\tau_c \simeq E^2/(2Knv)$. A further condition for the diffusive approximation
of transport to be valid is that the mean free path is small compared to the length of the loop: $\lambda \ll L_{loop}$.

\section{Electron flux spectrum}

In solar flare studies, the electron {\it flux} spectrum $F(E,\mu,z)$ (electrons~cm$^{-2}$~s$^{-1}$~keV$^{-1}$), differential in energy $E$~(keV), is normally used instead of the electron phase-space distribution function $f(v,\mu,z)$. Using the identity $F(E, \mu,z) \, dE = v \, f(v,\mu,z) \, dv$, we see that these quantities are related through $F(E,\mu,z) = f(v,\mu,z)/m_e$.  The continuity equation for the electron flux $F(E,\mu,z)$ thus follows simply by multiplying Equation~(\ref{Eq:fp}) by $1/m_e$.

HXR imaging observations typically are carried out over characteristic timescales of tens of seconds, which is much longer than the electron transport time $L_{Loop}/v$ \citep[e.g.,][]{2011SSRv..159..107H}. Therefore, we can safely ignore the temporal dependence $\partial/\partial t$ in Equation~(\ref{Eq:fp}) and write the resulting stationary transport equation in energy variables:
\begin{equation}\label{eq:kin_flux}
\mu \, \frac{\partial F(E,\mu, z)}{\partial z} \,  = \frac{\partial}{\partial E} \, \left(\frac{Kn(z)F(E,\mu,z)}{E}\right)+
\frac{\partial }{\partial \mu} \left ( \frac{D_{\mu\mu}}{v} \, \frac{\partial F(E,\mu,z)}{\partial \mu } \right ) +
H_0(E,\mu,z) \,\,\, ,
\end{equation}
where the source term $H_0(E,\mu,z)$ (electrons~cm$^{-3}$~s$^{-1}$~keV$^{-1}$)
allows for the local acceleration of electrons in the loop. The standard simplified geometry assumed
 is here so that electrons are accelerated near the apex of the loop $z=0$ and then can propagate
towards the chromosphere.

\subsection{Standard model of parallel transport in a collisional plasma}

In the standard model, electrons are assumed to propagate down the loop with collisional losses but
without being scattered at all. Moreover, in this model, the simplifying assumption  is often made
that the particles velocity is along $z$, which is the guiding field $\mathbf{B}_{0}$,
meaning than the electrons are thought to be all field aligned with zero pitch angle.
The electron continuity equation~(\ref{eq:kin_flux}) then becomes
\begin{equation}
{\partial F(E,z) \over \partial z} - {\partial \over \partial E} \left ( \frac{Kn(z)}{E} \, F(E,z) \right )=F_0(E) \, S(z) \,\,\, ,
\label{cont}
\end{equation}
where we have characterized the source of electrons by a separable form consisting of an injected
spectrum $F_0(E)$ (electrons~cm$^{-2}$~s$^{-1}$), spatially distributed throughout the source
according to the form of $S(z)$ (cm$^{-1}$).

To compare with spatially resolved X-ray observations we consider a source of energetic electrons (acceleration region)
with the (normalized) Gaussian spatial form
\begin{equation}\label{eq:S_z}
S(z)=\frac{1}{\sqrt{2\pi d^2}} \, \exp\left(-\frac{z^2}{2d^2}\right) \,\,\, ,
\end{equation}
where $d$ is the characteristic size of the acceleration region. Let us also assume that the source injects electrons
with a power-law energy spectrum
\begin{equation}\label{eq:F_0}
F_0(E)=\frac{\dot{N}}{A}\frac{(\delta -1)}{E_0}\left(\frac{E_0}{E}\right)^{\delta} \,\,\, , \,\,\,  E>E_0
\end{equation}
where $\delta $ is the electron spectral index and $E_0$ is the low energy cut-off. The electron flux spectrum is normalized
\begin{equation}\label{eq:F0}
\dot{N}=\int_{E_0}^{\infty}F_0(E)dE
\end{equation}
to the electron injection rate $\dot{N}$, which is the quantity that is deduced from the observation.

Let us first consider equation (\ref{cont}) with delta functions as the
source of particles in space when the particles injected parallel to $\mathbf{z}$, e.g. $\mu=1$
\begin{equation}
{\partial G_{+}(E,z) \over \partial z} - {\partial \over \partial E} \left ( \frac{Kn(z)}{E} \, G_{+}(E,z) \right )= F_0(E)\, \delta(z-z_0) \,\,\, ,
\label{cont_1}
\end{equation}
The solution of Equation (\ref{cont_1}) for $z>z_0$ so that $G_{+}(E, z=z_0+0)=F_0(E)$, $F(E, z=z_0-0)=0$  is
\begin{equation}
G_{+}(E,z;z_0)=\frac{E}{E_0}F_0(E_0)\theta(z-z_0)
\label{G_1}
\end{equation}
where $E_0^2(E,z;z_0)=E^2+2K \int _{z_0}^{z}n(z')dz'$ and $\theta(z)$ is the Heaviside step function.
$G_{+}(E,z;z_0)$ is the Green's function, so for an arbitrary source $S(z)$ of electrons, we find
\begin{equation}
F_{+}(E,z)=\int_{-\infty}^{\infty}G_{+}(E,z;z_0)S(z_0)dz_0
\label{F+}
\end{equation}
For the electrons moving antiparallel to $\mathbf{z}$ with $\mu=-1$, e.g. one can write
\begin{equation}
-{\partial G_{-}(E,z) \over \partial z} - {\partial \over \partial E} \left ( \frac{Kn(z)}{E} \, G_{-}(E,z) \right )= F_0(E)\, \delta(z-z_0) \,\,\, ,
\label{cont_2}
\end{equation}
with the solution
\begin{equation}
G_{-}(E,z;z_0)=\frac{E}{E_0}F_0(E_0)\theta(z_0-z)
\label{G_2}
\end{equation}
the corresponding solution becomes
\begin{equation}
F_{-}(E,z)=\int_{-\infty}^{+\infty}G_{-}(E,z;z_0)S(z_0)dz_0=\int_{-\infty}^{+\infty}\frac{E}{E_0}F_0(E_0)\theta(z_0-z)S(z_0)dz_0
\label{F-}
\end{equation}
where $E_0^2(E,z;z_0)=E^2+2K \int _{z}^{z_0}n(z')dz'$.
The solution of Equation (\ref{cont}) with electrons injected towards both footpoints
(e.g. $\mu=\pm 1$ )over $-\infty<z<+\infty$ can be written
\begin{equation}
F_{C}(E,z) =\frac{F_{-}+F_{+}}{2}=\frac{E}{2}\int_{-\infty}^{+\infty}\frac{F_0(E_0[E,z;z'])}{E_0[E,z;z']}
\, S(z') \, dz' \,\,\, ,
\label{Sol_F_Ex}
\end{equation}
where $E_0^2(E,z;z')=E^2+2K\mid \int _{z'}^{z}n(z'')dz'' \mid$. In the solution (\ref{Sol_F_Ex}), we have introduced a factor $1/2$
to account for the fact that the electrons propagate both ways, so the injection of electrons is double what is expected from the continuity
equation (\ref{cont}) but without a source and for the unidirectional particle transport in $0<z<+\infty$ that is often considered in transport models
for non-thermal electrons in solar flares \citep[see, e.g.,][]{1972SvA....16..273S}.
The collisional stopping distance $\lambda _{c}(E)$ is thus $\propto E^{2}$, a result that can be readily seen by simply
comparing the advective and energy loss terms: $F/\lambda _{c}(E) \sim KnF/E^2$, so that $\lambda _{c}(E) \sim E^2/Kn$.

In order to compare with spatially resolved HXR observations, the density-weighted mean electron flux must be calculated.
Multiplying by the local density $n(z)$ and integrating solution (\ref{Sol_F_Ex}) over the emitting volume, one finds

\begin{equation}
\langle nVF_{C}(E)\rangle \equiv\int_{V}F_{C}(E,z) \, n(z) \, dV=A \int_{-\infty}^{+\infty}F_{C}(E,z) \, n(z) \, dz=A \, \frac{E}{K}\int _{E}^{\infty} F_0(E') \, dE' \,\,\, ,
\label{eq:TT}
\end{equation}
where $A$ (cm$^2$) is the cross-sectional area of the loop. Equation ({\ref{eq:TT}}) is a standard expression for a thick-target density weighted
electron flux spectrum \citep[e.g.][]{1971SoPh...18..489B} and can be directly inferred from X-ray data \citep[e.g.,][]{2011SSRv..159..107H}.

Observationally, the density-weighted mean electron flux spectrum $\langle nVF(E)\rangle$ \citep[e.g.,][]{2003ApJ...595L.115B} can be readily deduced from the spatially-integrated hard X-ray spectrum $I(\varepsilon)$ [photons~cm$^{-2}$~s$^{-1}$~keV$^{-1}$] observed at the Earth:

\begin{equation}
I(\epsilon) = \frac{1}{4\pi R^2}\int _{\varepsilon}^{\infty}\langle nVF(E)\rangle \, \sigma(\varepsilon, E) \, dE \,\,\, ,
\label{eq:I_e}
\end{equation}
where $R$ is the Sun-Earth distance and $\sigma(\varepsilon, E)$ is the angle-averaged bremsstrahlung cross-section.  Equations~(\ref{eq:TT}) and~(\ref{eq:I_e}) show that observations of $I(\varepsilon)$ allows us to deduce the injection (i.e., acceleration) rate $A \, F_0(E)$ (electrons~s$^{-1}$~keV$^{-1}$). In practice, when compared with, e.g., {\em RHESSI} hard X-ray data, the accelerated electron spectrum is approximated by a power-law form $F_0(E_0) = C_0 \, E_0^{-\delta}$ (see Equation~(\ref{eq:F0})) and fitted to the data to find the best-fit parameters $(C_0,\delta)$.

\subsection{Diffusive transport in a collisional plasma}

Let us now consider the possibility that the magnetic loop is filled with plasma turbulence, so that as the particles propagate downwards, they experience pitch-angle scattering such that the angular distribution of energetic electrons becomes isotropic on a scale $\lambda << L_{Loop}$.  In this case,
the collisional transport model, Equation (\ref{cont}), becomes

\begin{equation}\label{eq:kin_flux_diff}
\frac{1}{v} \, \frac{\partial }{\partial z}\left(D^{(T)}_{zz} \, \frac{\partial F}{\partial z}\right) \, = \,
\frac{\partial}{\partial E}\left(\frac{Kn(z)}{E} \, F\right) \, + \, F_0(E) \, S(z) \,\,\, ,
\end{equation}
where the advective term has been replaced by the diffusive term.

Assuming a uniform density $n(z)=n_0$, Equation~(\ref{eq:kin_flux_diff}) can be solved analytically using a Green's function approach. Following \citet[][]{1959SvA.....3...22S}, we first solve Equation~(\ref{eq:kin_flux_diff}) for the electron flux spectrum $G(E,z)$ corresponding to a point source of monoenergetic electrons $F_0(E) \, S(z)=\delta(E-E') \, \delta(z-z')$.  Dividing by $Kn_0$ and using the form of $D^{(T)}_{zz}$ from Equation~(\ref{eq:D_zz1}), Equation (\ref{eq:kin_flux_diff}) reads

\begin{equation}\label{eq:kin_flux_diff_G}
\frac{\lambda}{3Kn_0} \, \frac{\partial^2 G}{\partial z^2} =
\frac{\partial}{\partial E}\left(\frac{G}{E}\right) +\frac{1}{Kn_0} \, \delta(z-z') \, \delta(E-E') \,\,\, .
\end{equation}
This can be further simplified by changing variables $\xi =E^2$ and $B=G/E$:

\begin{equation}\label{eq:kin_flux_diff_B}
a \, \frac{\partial ^2 B}{\partial z^2} \, - \,
\frac{\partial B}{\partial \xi} =\frac{1}{Kn_0} \, \delta(z-z') \, \delta(\xi-\xi') \,\,\, ,
\end{equation}
where $a=\lambda/(6Kn_0)$. Equation~(\ref{eq:kin_flux_diff_B}) is a standard diffusion equation, which has the solution, valid in  $-\infty<z<\infty$ and $\xi -\xi ' >0$,

\begin{equation}\label{eq:B_solution}
B(\xi,z;\xi',z')=\frac{1}{Kn_0} \, \frac{1}{\sqrt{4\pi a(\xi-\xi ')}} \, \exp\left(-\frac{(z-z')^2}{4 a (\xi-\xi')}\right) \, \theta(\xi-\xi') \,\,\, ,
\end{equation}
where $\theta(x)$ is the Heaviside step function, so that $d\theta(x)/dx=\delta(x)$.

Using the Green's function solution (\ref{eq:B_solution}), one readily finds by superposition the solution $F_D (E,z)$ of Equation (\ref{eq:kin_flux_diff}) for an arbitrary injection flux spectrum $F_0(E)$ and arbitrary spatial injection distribution $S(z)$:

\begin{equation}\label{eq:F_solutionD}
F_{D}(E,z)=\frac{E}{Kn_0}\int _{-\infty}^{\infty}dz' \int _{E}^{\infty}dE'
\frac{F_0(E') \, S(z')}{\sqrt{4\pi a(E'^2-E^2)}} \, \exp\left(-\frac{(z-z')^2}{4 a(E'^2-E^2)}\right) \,\,\, .
\end{equation}

That the diffusional stopping distance $L \propto a^{1/2} E \propto (\lambda/Kn_0)^{1/2} \, E$ is readily seen from the form of the exponential term in Equation~(\ref{eq:F_solutionD}).  This result can also be found simply by balancing the diffusion term with the collisional term in Equation~(\ref{eq:kin_flux_diff}). This leads to $D_{zz}/vL^2 \sim Kn/E^2$, so
that $L \propto \sqrt{D_{zz} E^{3/2}/Kn_0}$. Since $D_{zz} \propto \lambda \, v \propto \lambda \, E^{1/2}$,
$L \propto (\lambda/Kn_0)^{1/2} \, E$.

Similarly to the collisional transport case, the solution  (\ref{eq:F_solutionD}) can be integrated to find the density-weighted
spatially-integrated spectrum (i.e., mean electron flux)

\begin{equation}\label{eq:VnF_solutionD}
\langle nVF_D(E) \rangle=\int _{V}F_D(E,z) \, n_0 \, dV =A \, n_0 \, \int_{-\infty}^{\infty}F_{D}(E,z) \, dz=\frac{E}{K}\int _{E}^{\infty} A \, F_0(E') \, dE' \,\,\, ,
\end{equation}
where the last equality follows from changing the order of integration after substituting for $F_D(E,z)$ from Equation~(\ref{eq:F_solutionD}). {\it The spatially integrated mean flux $\langle nVF_D(E)\rangle $ is exactly the same as the spatially integrated flux given by the collisional transport equation (\ref{eq:TT}).} This simply reflects the fact\footnote{If $F(E,z)\rightarrow 0$ at $z \rightarrow \pm \infty$, i.e. the particles lose their energy within finite distance, then the transport terms $\partial F/\partial z$ in Equation (\ref{Sol_F_Ex}) or the diffusive transport term $\partial ^2 F/\partial z^2$ in Equation (\ref{eq:kin_flux_diff}) becomes zero at at $z \rightarrow \pm \infty$ and the spatially integrated flux spectrum is independent of the form of the spatial and pitch angle evolution of the electrons. Therefore, the integration always leads to the expression~(\ref{eq:VnF_solutionD}). } that the total emitted flux in a thick target is independent of the details of the pitch angle evolution.

\section{Spatial distribution of energetic electrons and hard X-ray emission in a diffusive transport model}

The spatial distribution of energetic electrons along the magnetic loop can be found from Equations~(\ref{eq:F_solutionD}) and~(\ref{eq:S_z}):

\begin{equation}\label{eq:F_solutionS}
F_D(E,z)=\frac{E}{Kn_0}\int _{E}^{\infty}dE'
\frac{F_0(E')}{\sqrt{4\pi a(E'^2-E^2)+2d^2}}\exp\left(-\frac{z^2}{4a(E'^2-E^2)+2d^2}\right) \,\,\, .
\end{equation}
For comparison we can also write the solution for the standard collisional transport case using equations (\ref{eq:S_z}) and (\ref{eq:F0})

\begin{equation}\label{eq:F_solutionF}
F_C(E,z) =\frac{E}{2}\int_{-\infty}^{+\infty}\frac{F_0(E_0[E,z;z'])}{E_0[E,z;z']}
\, \frac{1}{\sqrt{2\pi d^2}} \, \exp\left(-\frac{z'^2}{2d^2}\right)  \, dz' \,\,\, ,
\end{equation}
where $E^2_0[E,z;z']=E^2+2Kn_0\mid{z-z'}\mid$.

\begin{figure}[pht]
\begin{center}
\includegraphics[width=89mm]{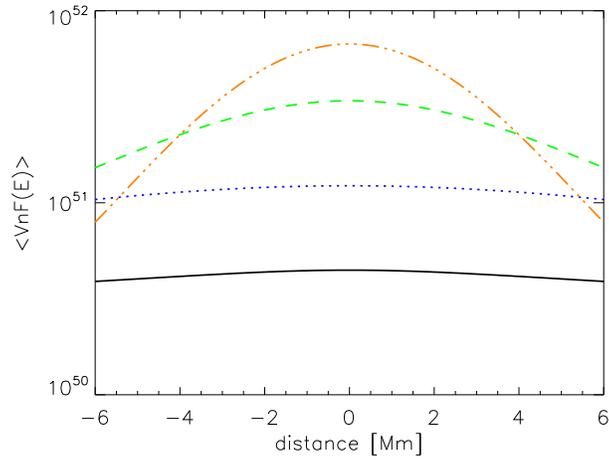}
\end{center}
\caption{Spatial distribution of energetic electrons $A \, \Delta z \, F(E,z)$ at 20~keV, for a density $n_0=10^{10}$~cm$^{-3}$
and an injection (acceleration) rate $\dot{N}=10^{36}$~s$^{-1}$ above $E_0=10$~keV, with $A\, \Delta z=10^{26}$~cm$^{-3}$
and $\delta =4$, $d=2$~Mm. The standard transport case is shown by a black solid line and the diffusive transport cases are shown for $\lambda =10^{9}$~cm (blue dotted line), $\lambda =10^{8}$~cm (green dashed line), and $\lambda =10^{7}$~cm (orange dot-dashed line).}
\label{fig:cor_source_distribution}
\end{figure}

\begin{figure}[pht]
\begin{center}
\includegraphics[width=89mm]{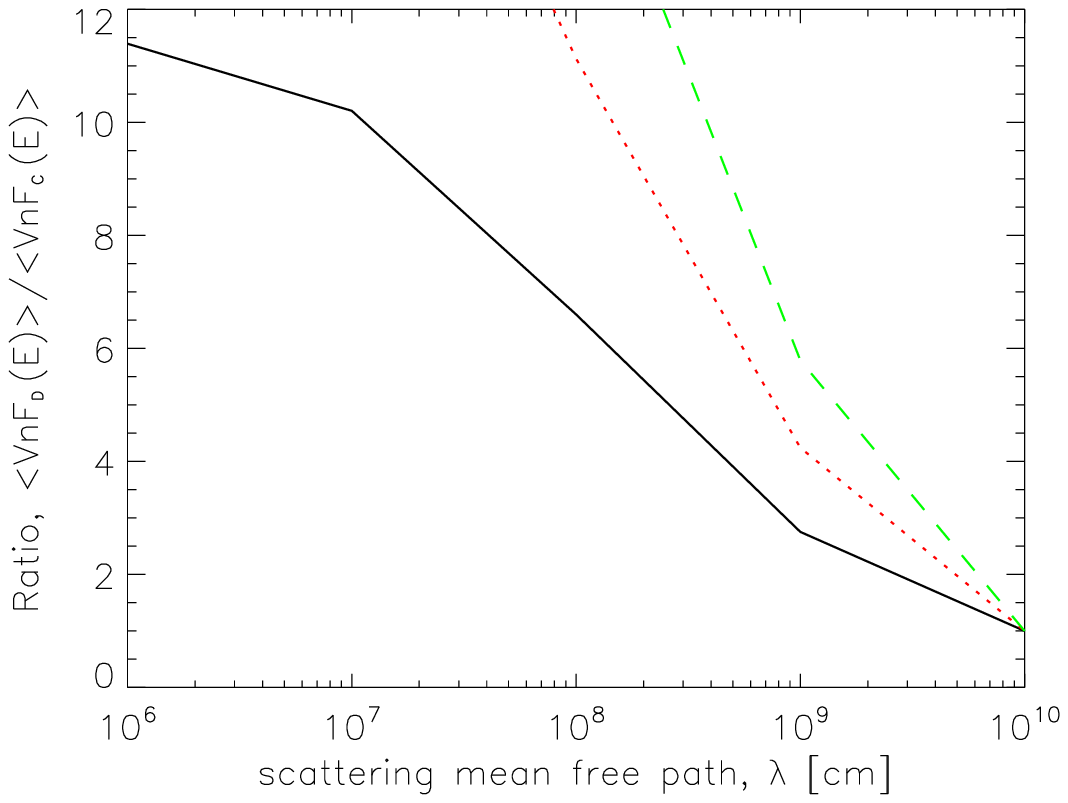}\\
\includegraphics[width=89mm]{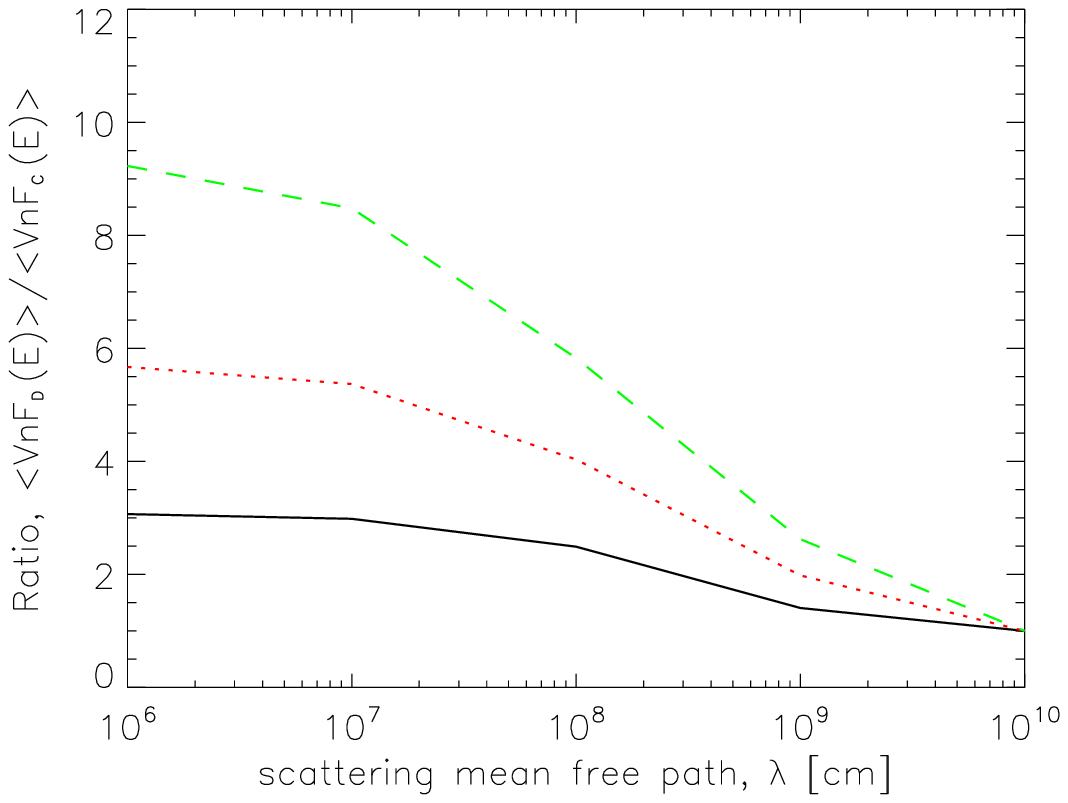}\\
\includegraphics[width=89mm]{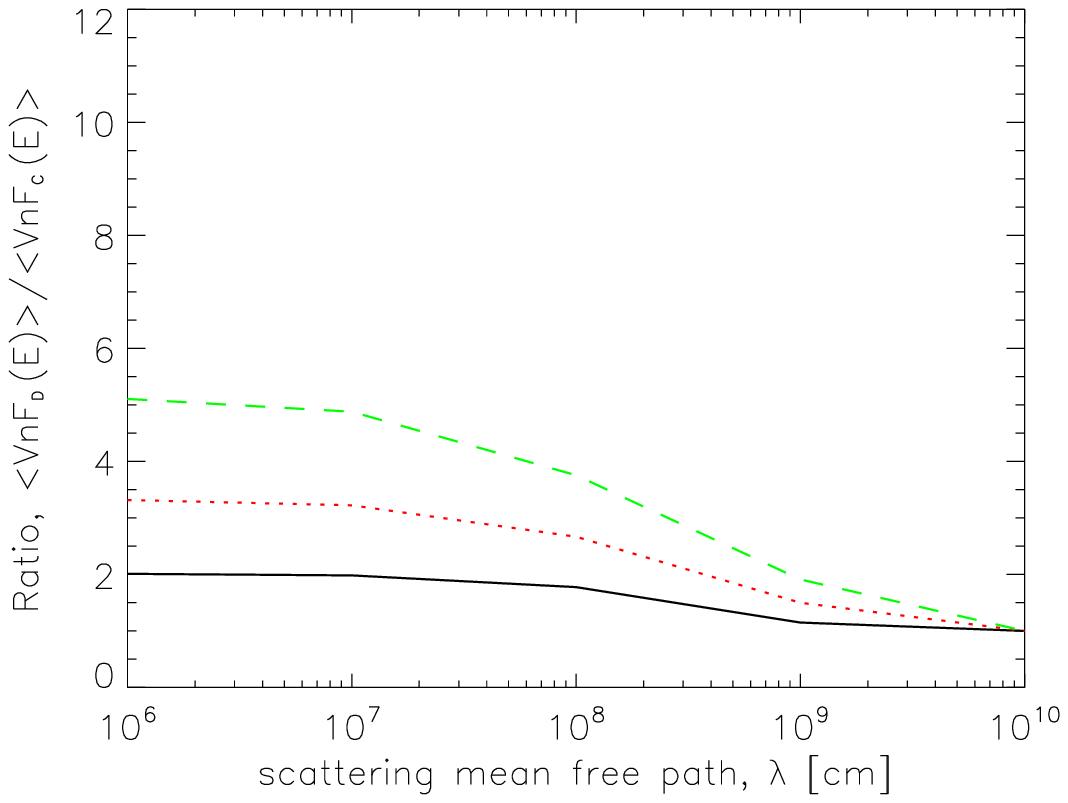}
\end{center}
\caption{Ratio of the mean electron fluxes $\langle nVF_D(E)\rangle/\langle nVF_C(E)\rangle$ in the coronal source defined by Equation (\ref{eq:F_CS})
for plasma densities: $n_0=1\times10^{10}$~cm$^{-3}$ (top), $n_0=5\times10^{10}$~cm$^{-3}$ (middle), and $n_0=1\times10^{11}$~cm$^{-3}$ (bottom). Three characteristic energies are considered: 20~keV (solid black line), 30~keV (orange dashed line), and 40~keV (red dashed line).}
\label{fig:ratios}
\end{figure}

The solutions for the diffusive (\ref{eq:F_solutionS}) and streaming (\ref{eq:F_solutionF}) cases are compared in Figure~\ref{fig:cor_source_distribution}, for typical flare parameters.  Pitch angle scattering causes electrons to escape the acceleration region more slowly, which results in an enhanced electron number in the coronal source (Figure~\ref{fig:ratios}). As an example, for a loop density $n_0=10^{10}$~cm$^{-3}$ and a mean free path $\lambda =10^6$~cm, the electron flux $F_D(E,z)$ is greater than that for the standard transport case $F_C(E,z)$ by a factor of $\sim$20. The shorter the mean free path due to non-collisional scattering, the stronger the enhancement.
As the mean free path $\lambda \rightarrow 0$, the coronal source effectively becomes a ``thick-target'' source.  Although the density is not high enough  to collisionally stop the electrons, the efficient scattering of electrons leads to effective electron trapping, so that the electrons lose most of their energy within the coronal part of the loop.

\begin{figure}[pht]
\begin{center}
\includegraphics[width=89mm]{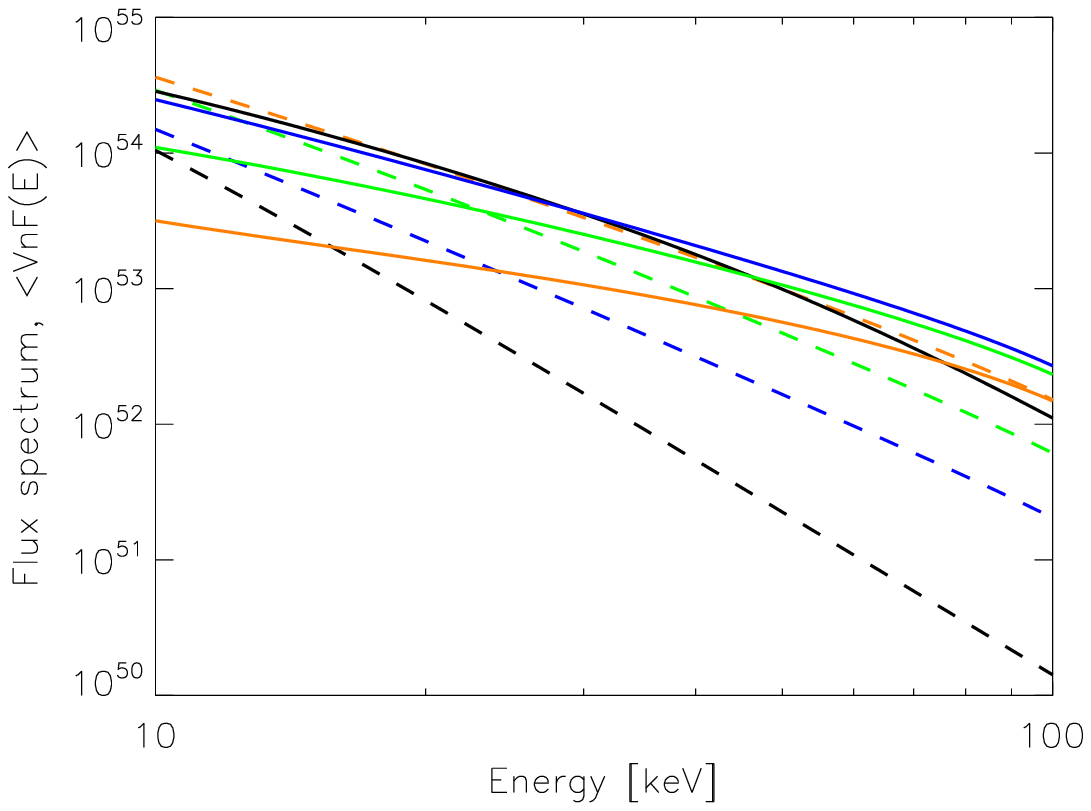}
\includegraphics[width=89mm]{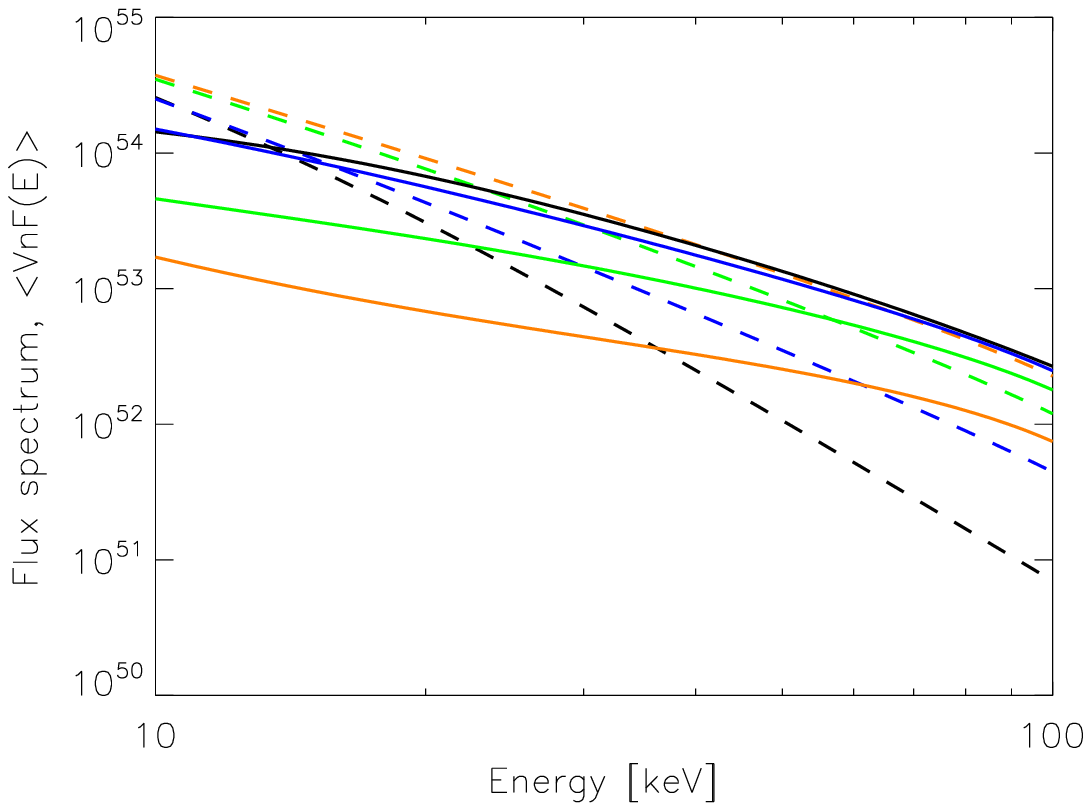}
\includegraphics[width=89mm]{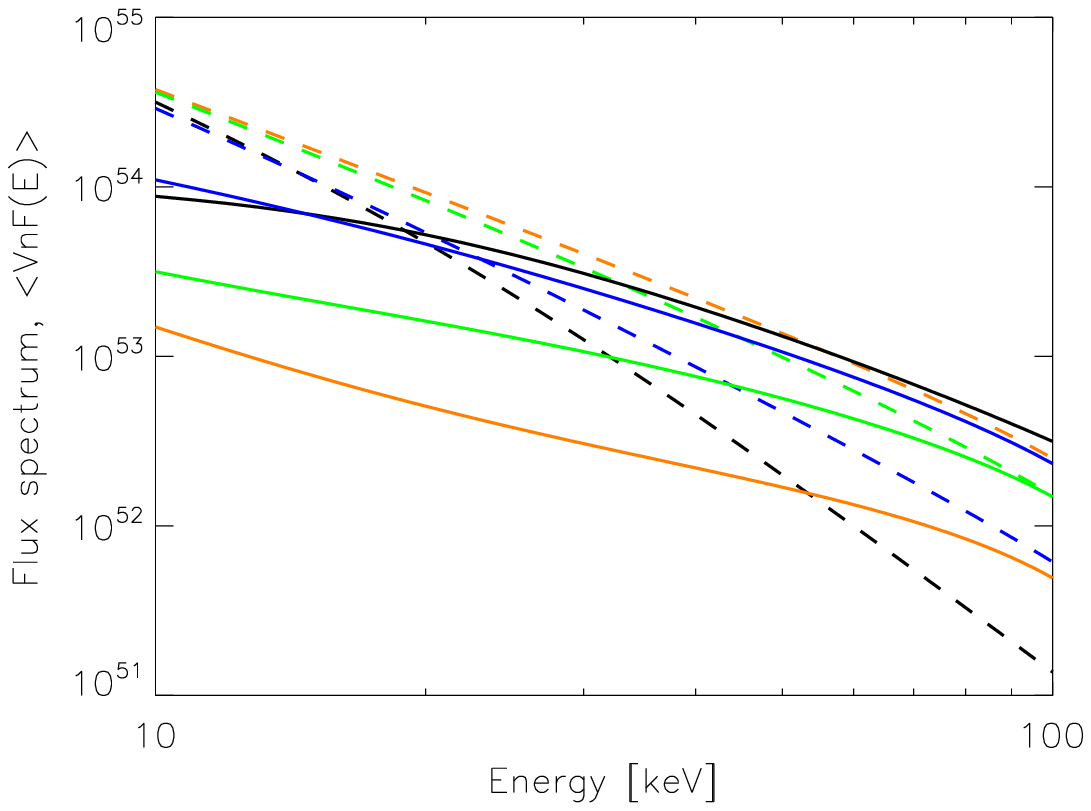}
\end{center}
\caption {Mean electron flux spectrum $\langle nVF(E)\rangle$ for three plasma densities $n_0=1\times 10^{10}$~cm$^{-3}$ (top panel), $n_0=5\times 10^{10}$~cm$^{-3}$ (middle panel), $n_0=1\times 10^{11}$~cm$^{-3}$ (bottom panel). The dashed lines show the spectrum of the coronal source $\langle nVF^{CS}(E)\rangle$ and the solid black lines show the spectrum of the footpoints $\langle nVF^{FP}(E)\rangle$. Four transport cases are shown: scatter-free (black lines), diffusive with $\lambda=10^9$~cm (blue lines), diffusive with $\lambda=10^8$~cm (green lines), and diffusive with $\lambda=10^7$~cm (orange lines).}
\label{fig:mean_flux}
\end{figure}

We can compare the intensities of emission from the foot-point and coronal sources.  Define the coronal emission as

\begin{equation}\label{eq:F_CS}
\langle nVF^{CS}(E)\rangle =An_0\int _{-HWFM}^{+HWFM} F(E,z) \, dz \,\,\, ,
\end{equation}
where $HWHM=\sqrt{2\ln{2}} \, d$ is the half-width at half-maximum.
Similarly, the footpoint emission is defined as

\begin{equation}\label{eq:F_FP}
\langle nVF^{FP}(E)\rangle =2An_0\int_{HWHM}^{\infty} F(E,z) \, dz \,\,\, .
\end{equation}
The sum of the two sources (\ref{eq:F_CS}) and (\ref{eq:F_FP}) is, of course,

\begin{equation}\label{eq:F_TT}
\langle nVF^{FP}(E)\rangle+\langle nVF^{CS}(E)\rangle=\frac{E}{K}\int _{E}^{\infty} A \, F_0(E') \, dE' \,\,\, ,
\end{equation}
the spatially-integrated flux spectrum. The solutions presented in Figure~\ref{fig:mean_flux} for three typical plasma densities (and for an electron spectral index $\delta =4$) allow comparison with {\em RHESSI} imaging-spectroscopy observations.

The influence of pitch angle scattering is evidenced by stronger coronal emission and weaker foot-point emission than in the standard case due to the increase of the residence time of electrons high-up in the corona. Turbulent pitch angle scattering also leads to a change in the HXR spectral index, forming a broken-power-law spectrum, a feature noticed by \citet{1991ApJ...374..369B}. For collisional transport in a medium of density $n_0=1\times 10^{10}$~cm$^{-3}$ (see top panel of Figure~\ref{fig:mean_flux}), the coronal source has a spectrum $\langle nVF^{CS}(E)\rangle \propto E^{-4}$ and the footpoint spectrum $\langle nVF^{FP}(E)\rangle \propto E^{-2}$. In the diffusive cases, the coronal emission becomes stronger, and the spectrum progressively flatter, with decreasing $\lambda$, while the footpoint spectrum develops a break and becomes weaker at energies in the low tens of keV.  The effect of enhanced electron density in the coronal part of the loop is stronger at low energies, despite the fact that the pitch angle scattering rate grows with speed according to $D_{\mu \mu}^{(T)} \propto v/\lambda$. This is related to the fact that the solution $F_D(E,z)$ given by (\ref{eq:F_solutionS}) depends on the ratio $\lambda/\lambda_c(E)$, where $\lambda _c(E)=E^2/2Kn_0$ is the collisional stopping depth of electrons of energy $E$. For large energies $\lambda/\lambda_c(E)$ is smaller, so the electrons with $E^2>2Kn_0\lambda$ (i.e., $\lambda/\lambda_c < 1$) can reach the footpoints.

\begin{figure}[pht]
\begin{center}
\includegraphics[width=69mm]{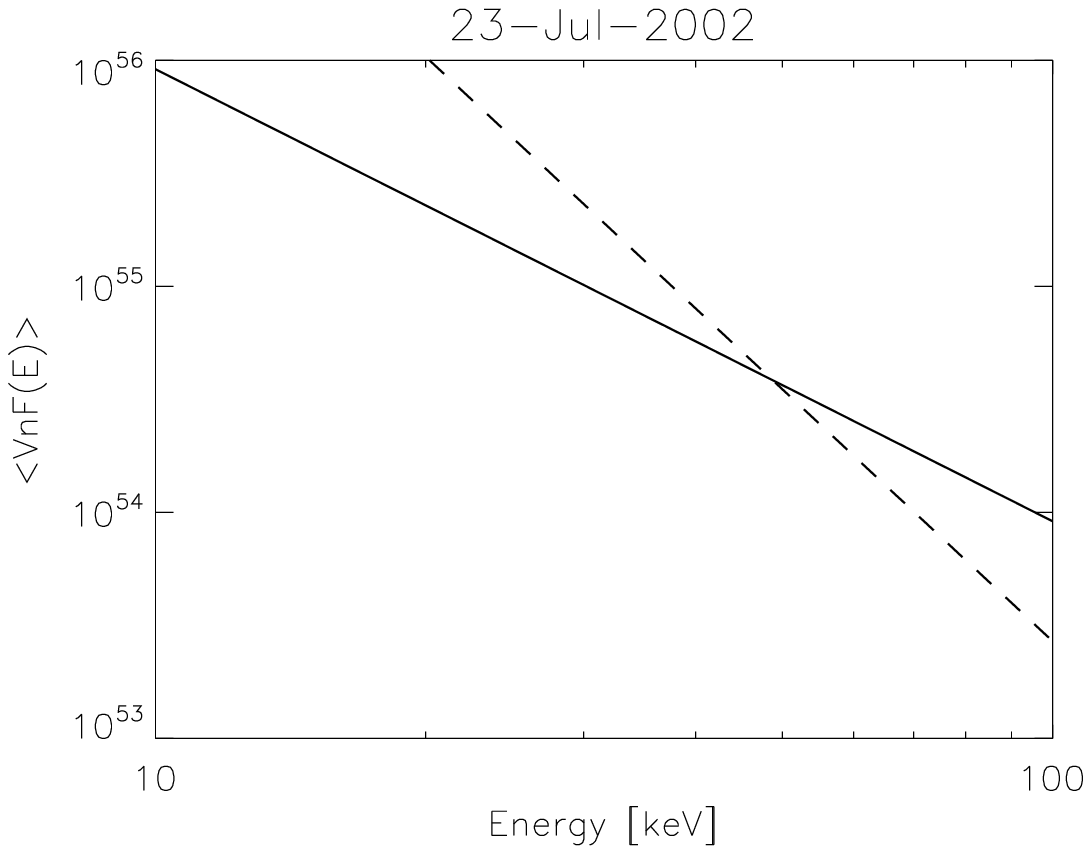}
\includegraphics[width=69mm]{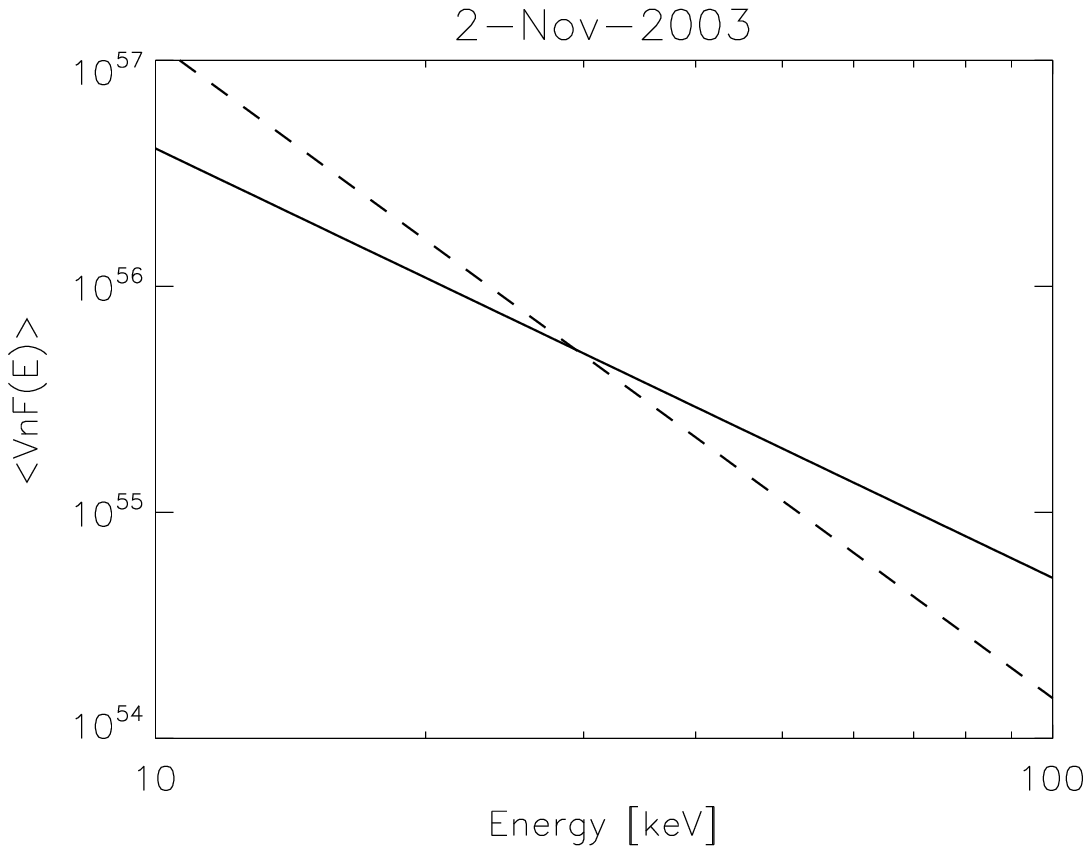}\\
\includegraphics[width=69mm]{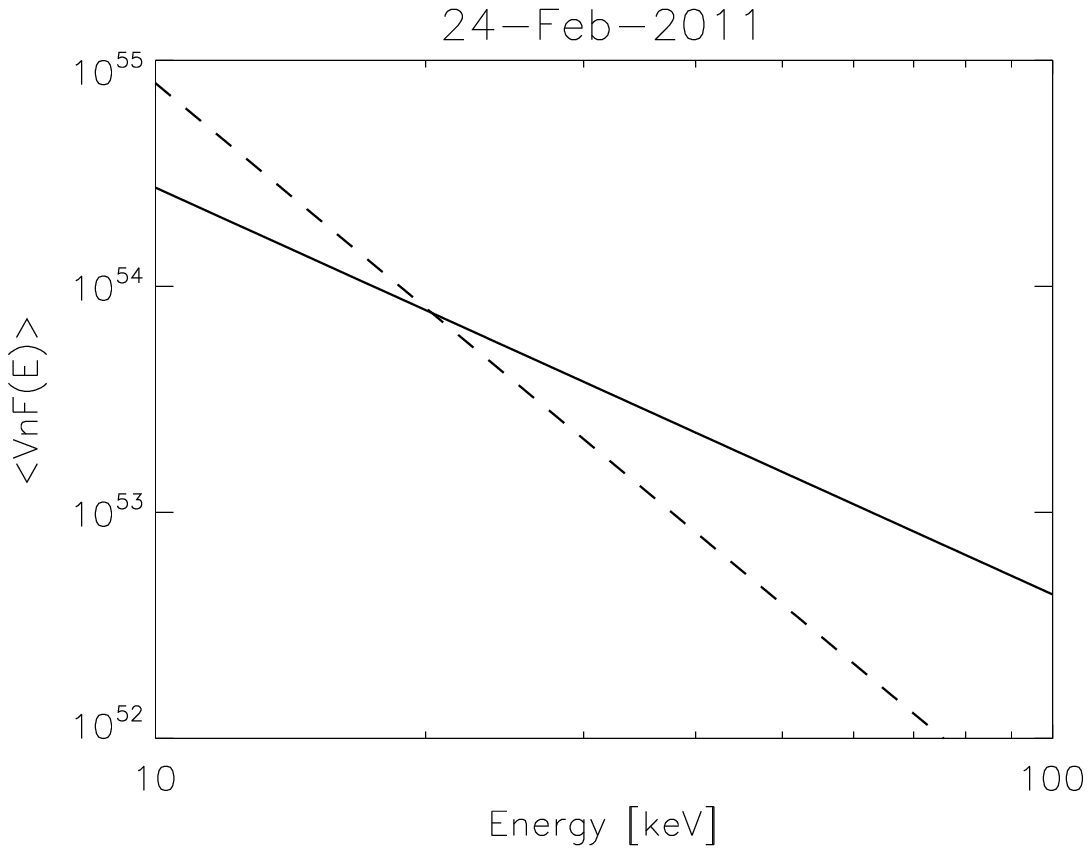}
\includegraphics[width=69mm]{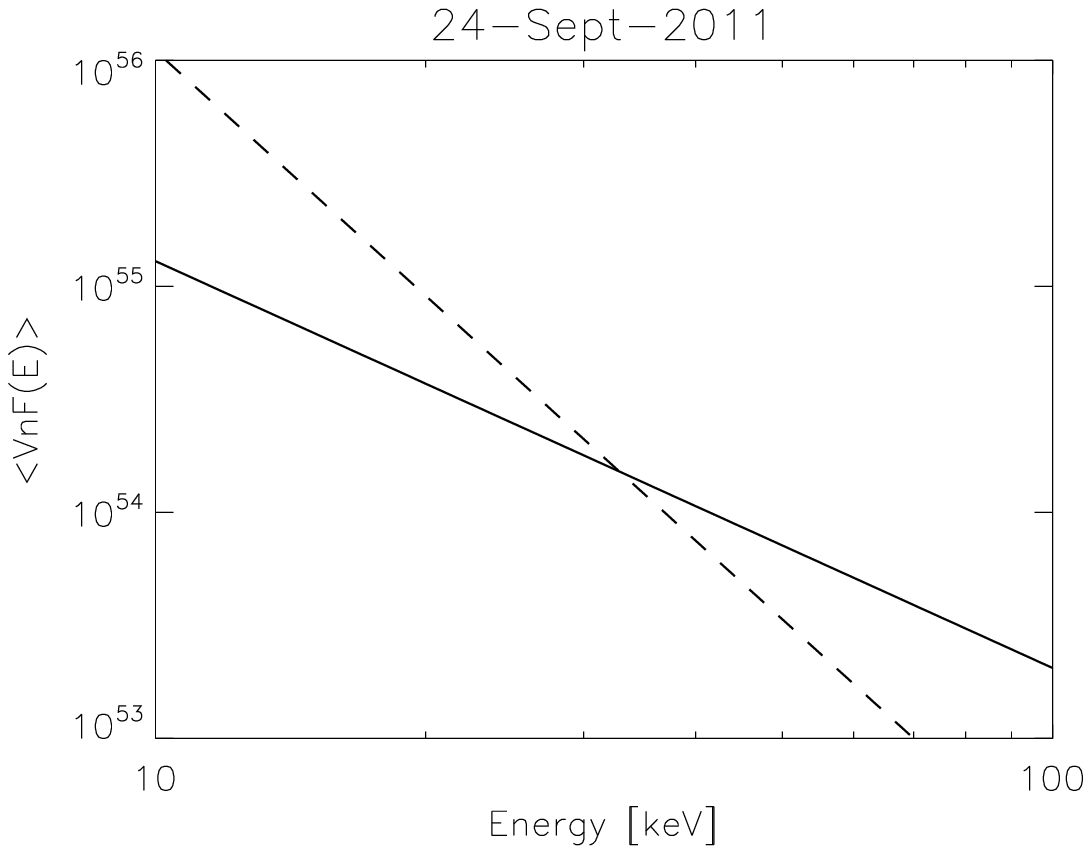}
\end{center}
\caption{The coronal source (dashed line) and footpoint (solid line) mean electron fluxes $\langle nVF(E)\rangle$ for four flares analyzed by \citet[][]{2013A&A...551A.135S}. The figure shows power-law fits to imaging spectroscopy results. The typical uncertainties on the spectral index are $\pm 0.2$ and on the mean electron flux $\pm 20$\%. As {\em RHESSI} has limited dynamic range, the most reliable range of energies is where the fluxes are comparable.}
\label{fig:flares}
\end{figure}

Recent {\em RHESSI} observations by \citet{2013A&A...551A.135S} of four well-resolved flares with both coronal and foot-point sources suggest (see Figure~\ref{fig:flares}) that the number of electrons in the coronal part of the loop is larger by a significant factor (between $\sim$2 and $\sim$8) than what is required to explain the thick-target footpoint emission. The likely source of this discrepancy is the trapping of energetic electrons in the solar corona, probably in the acceleration region itself. Figure~\ref{fig:ratios} shows the enhancement of flux spectrum in the coronal source for various plasma densities and scattering mean free paths $\lambda$. For example, in the flare of 2011~February~24, the flaring loop density was $n_0 \sim 5\times 10^{10}$~cm$^{-3}$ and the electron spectral index $\delta=4$, as deduced from {\em RHESSI} observations. To obtain the intersection between the coronal $\langle nVF^{CS}(E)\rangle$ and foot-point $\langle nVF^{FP}(E)\rangle$ spectra near 20~keV, as required by {\em RHESSI} observations (Figure~\ref{fig:flares}), the non-collisional mean free path $\lambda$ should not be less than a few thousand km. The green lines (both solid and dashed) in the middle panel of Figure~\ref{fig:mean_flux} shows that for $\lambda =10^8$~cm the coronal source will dominate up to around 50~keV, which is inconsistent with the observations. Analysis of other events analyzed by \citet{2013A&A...551A.135S} and presented in Figure~\ref{fig:flares} suggest that the scattering mean free path $\lambda$ must be of order $(10^8-10^9)$~cm. We notice that smaller $\lambda$ (e.g., $\lambda <10^8$~cm) will noticeably reduce the foot-point HXR emission (see Figure~\ref{fig:mean_flux}) to an extent that the ratio of intensities of the coronal and footpoint sources would be inconsistent with the {\em RHESSI} data.

\subsection{Dependence of coronal source size on energy}

Spatially resolved observations of HXR loops at various energies provide additional constraints on the poorly-known level of magnetic fluctuations in solar flare loops and allow us to derive the pitch angle scattering length. Recent observations \citep{2008ApJ...673..576X,2011ApJ...730L..22K,2012A&A...543A..53G}
suggest that the length of HXR coronal sources is energy-dependent, with the FWHM length of the loop growing as $L\simeq L_0+\alpha E^2$, where $L_0$ is the characteristic length of the acceleration (injection) region and $\alpha$ is a coefficient that is generally consistent with collisional transport, i.e., $\alpha \simeq 1/(2Kn)$. In addition, the FWHM {\it width} of coronal loops grows slowly with energy \citep{2011ApJ...730L..22K}, which is consistent with electron transport in a fluctuating magnetic field. The form $L\simeq L_0+ E^2/(2Kn)$ directly follows from collisional transport without scattering (i.e., from Equation~(\ref{eq:F_solutionF})). However, in the diffusive transport model,

\begin{equation}\label{Ldiffusive}
L(E)-L_0 \propto \lambda^{1/2} E \,\,\, ,
\end{equation}
i.e., the source size grows {\it linearly} with energy, with slope proportional to $\lambda^{1/2}$ (see Equation~(\ref{eq:F_solutionD}) and remarks thereafter). In the limit of strong diffusion ($\lambda \rightarrow 0$), the source size will be essentially independent of energy. The diffusive solution~(\ref{eq:F_solutionS}) therefore allows us estimate the range of $\lambda$ which could be consistent with the observations.

As an example, we consider the well-studied 2002~April~15 flare, previously analyzed by \citet{2008ApJ...673..576X}, \citet{2011ApJ...730L..22K}, and \citet{2012A&A...543A..53G}.  This flare is characterized by a high plasma density around $n_0=2\times 10^{11}$~cm$^{-3}$, so that X-ray producing electrons up to around 30~keV are collisionally stopped within the coronal part of the loop.

\begin{figure}[pht]
\begin{center}
\includegraphics[width=89mm]{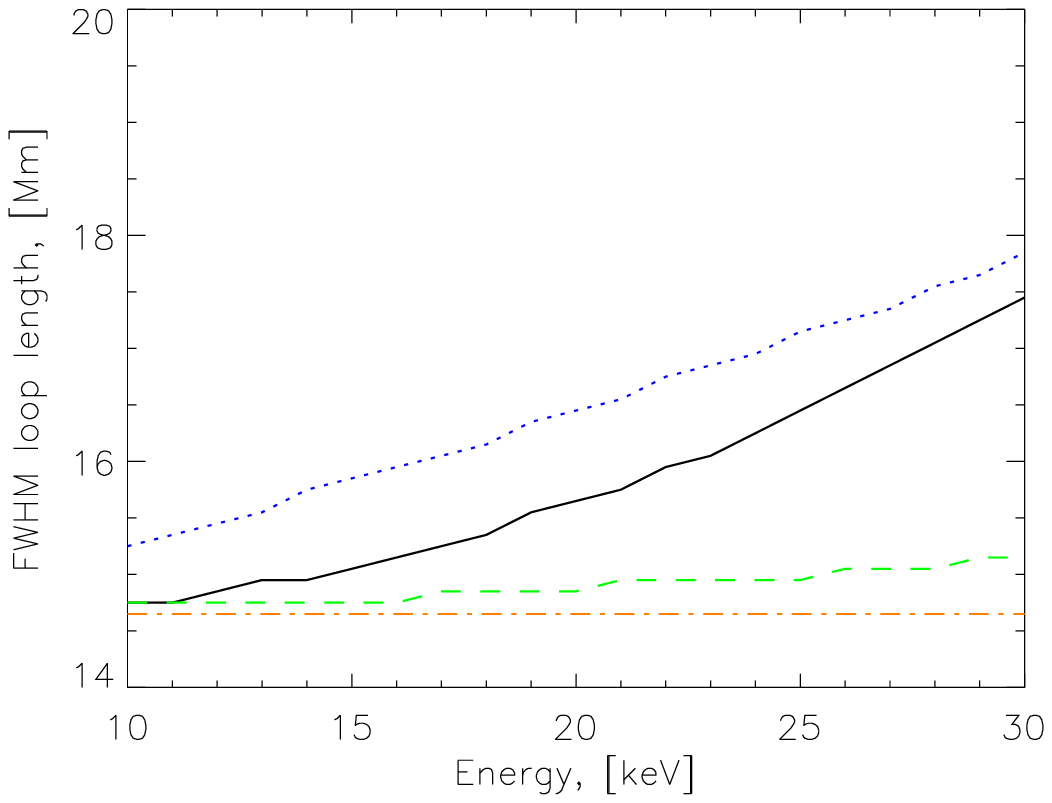}
\end{center}
\caption{Predicted FWHM length of the source as a function of energy in a loop with density $n_0=2\times 10^{11}$~cm$^{-3}$.
The electron spectral index $\delta =7$ and the acceleration/injection
region FWHM$ =2d \, \sqrt{2\ln 2}\simeq2.35 \, d\simeq14.5$~Mm ($=20$~arcseconds), so that $d=6.2$~Mm,
similar to the values in \citet{2008ApJ...673..576X}, \citet{2011ApJ...730L..22K}, and \citet{2012A&A...543A..53G}.
The collisional transport case is shown by the black solid line (Equation \ref{eq:F_solutionF}). Diffusive transport
cases (calculated using Equation \ref{eq:F_solutionS}) with $\lambda =10^{9}$~cm (blue line),
$\lambda =10^{8}$~cm (green line), and $\lambda =10^{7}$~cm (orange line) are also shown.}
\label{fig:FWHM_length}
\end{figure}

Figure~\ref{fig:FWHM_length} shows the FWHM length of the electron source and the $E^2$ dependence of source length with energy appropriate to the standard transport case. This $E^2$ dependence becomes the linear dependence $\lambda^{1/2}E$ predicted by a diffusive model~(\ref{Ldiffusive}) for $\lambda =10^9$~cm, and by $\lambda =10^7$~cm, the length is essentially energy-independent. Preliminary analysis suggests that the uncertainties in the {\em RHESSI} observations do support a linear relationship between $L$ and $E$ (with an appropriately large value of $\lambda$), and we intend to perform a more detailed observational test of the predictions of the diffusive transport model in a future work.  However, we can nevertheless conclude from the fact that there is a significant variation of $L$ with $E$ that very strong pitch angle scattering, e.g., $\lambda \lapprox 10^8$~cm, is {\it not} consistent with the observations \citep{2012A&A...543A..53G}.

\section{Summary and discussion}

We have considered the evolution of the electron flux spectrum $F(E,z)$ in a collisional plasma that contains a homogeneous distribution of magnetic fluctuations. The presence of these magnetic fluctuations leads to pitch-angle scattering of the hard X-ray producing electrons. In the approximation of strong pitch angle scattering over the size of the loop, this manifests itself as a diffusion parallel to the guiding magnetic field.  We have derived simple analytical solutions that allow us to compare {\em RHESSI} observations with this model in order to deduce limits on the mean free path associated with the scattering.

One of the interesting aspects of diffusive transport is the reduction of the direct current associated with the precipitating particles. The current density
in case of scatter-free propagation is $j_{C}\simeq e\dot{N}$, while the presence of pitch angle scattering will reduce this value
to $j_{D}\simeq e\dot{N}\lambda/(3L_{loop})$, so that
\begin{equation}\label{eq:current}
\frac{j_{D}}{j_C}\simeq \frac {\lambda}{3 L_{loop}}\,\,\, .
\end{equation}
As the return current \citep[e.g.][]{1980ApJ...235.1055E,2006ApJ...651..553Z} and associated ohmic losses are related to the direct current, these
will be reduced due to the non-collisional pitch angle scattering.

The non-collisional pitch angle scattering of electrons in the presence of collisional losses
makes the electron spectrum of the coronal source harder at low energies. In general, for the typical solar parameters
the coronal and footpoint spectra will be no longer single power-laws, but broken power-laws \citep[see also][]{1991ApJ...374..369B}.
Thus single power-law fits to the coronal and footpoint sources could lead to the spectral index differences not equal to 2.
We note that for the standard transport model, the spectral index difference between the coronal source
and foot-points is expected to be 2. Thus, the consideration of {\it non-collisional} pitch angle scattering can explain the
spread of the spectral index differences in spectral indices between the coronal and foot-point sources  observed in solar flares
\citep[e.g.][]{2003ApJ...595L.107E,2006A&A...456..751B}. While in this paper the scattering centers are assumed to be distributed
uniformly throughout the source, this may not be the case in an actual flare, and such an inhomogeneity could contribute
to the asymmetry  of foot-point spectral indices \citep{2008SoPh..250...53S}.

The analysis of spatially-resolved mean electron flux spectra in flares \citep{2013A&A...551A.135S} also suggests the presence of some trapping or pitch angle scattering in the coronal part of the loop, where the electrons are likely to be accelerated. The number of energetic electrons in the coronal source
exceeds the number required to explain foot-point emission. This can be seen from the graphs of mean electron spectra (Figure \ref{fig:flares}),
the flatter footpoint spectra tend to intersect with steeper coronal source spectra at higher energies than predicted by purely collisional transport (Figure \ref{fig:mean_flux}). Comparing  Figures (\ref{fig:mean_flux}) and (\ref{fig:flares}), one sees that the typical energies of intersection are better
explained with $\lambda$ in the range $\sim 10^8-10^9$~cm, which is shorter than the length of the loop.

For high loop densities \citep[e.g.,][]{2012A&A...543A..53G}, the variation of the FWHM of the X-ray source length with electron energy $E$ can be explained by collisional transport along the field lines.  However, it can also be explained by our collisional-diffusive model if the equivalent mean free path is comparable to the observed extent if the source. However, the mean free path cannot be smaller than about $\sim$10$^9$~cm, otherwise the predicted energy dependence of the source length would be too weak to be consistent with observations.

The inferred values of $\lambda$ are less than the typical length of a loop $\sim$2$\times 10^9$~cm, yet are comparable with the typical size of a coronal source $\sim$5$\times 10^8$~cm. These findings put constraints on the likely acceleration scenario inside a flaring loop. A scattering mean free path as large as the acceleration region requires that the acceleration itself does not rely on strong pitch angle scattering of deka-keV electrons.

The accumulation of electrons in the coronal source could also, in principle, be achieved via magnetic mirror trapping; however, the mirroring points must be inside the coronal sources in order to be consistent with the observations. This is rather atypical scenario for a simple loop geometry, in which the magnetic reflection points are normally near/at the footpoints, where the magnetic field strength significantly increases.  We further note that in a simple mirroring model the magnetic mirror points are determined only by the electron pitch angle and are thus energy independent, while the observations of high density loops strongly suggest sizes that are energy dependent. Therefore, in order for magnetic mirroring to be the chief trapping mechanism, one needs additional assumptions on the relation between the energy and pitch-angle distributions of the accelerated electrons, so that the higher energy ones could mirror back further from injection/acceleration site. While such a scenario cannot be ruled out completely, it does require further detailed numerical modeling to make quantitative statements.

The presented analysis of these flare suggest that the non-collisional pitch angle is likely to be present in solar flare loops, however the characteristic mean free path against this turbulent scattering is longer than $10^8-10^9$~cm (with some variation from flare to flare) and the characteristic time scale
is longer than $\sim \lambda/v=10^{-2}-10^{-1}$~s for $\sim 30$~keV (e.g. $v=10^{10}$~cm/s) electrons.

\acknowledgments
This work is supported by the STFC grant (E.P.K., N.H.B.). Financial support by the European Commission through the FP7 HESPE network (FP7-2010-SPACE-263086) is gratefully acknowledged; AGE was supported by NASA Grant NNX10AT78J. The authors are thankful to N. Jeffrey for helping to improve the text of the paper.

\bibliographystyle{apj}
\bibliography{refs_rhessi}

\begin{thebibliography}{54}
\expandafter\ifx\csname natexlab\endcsname\relax\def\natexlab#1{#1}\fi

\bibitem[{{Arnoldy} {et~al.}(1968){Arnoldy}, {Kane}, \&
  {Winckler}}]{1968ApJ...151..711A}
{Arnoldy}, R.~L., {Kane}, S.~R., \& {Winckler}, J.~R. 1968, \apj, 151, 711

\bibitem[{{Aschwanden} {et~al.}(2002){Aschwanden}, {Brown}, \&
  {Kontar}}]{2002SoPh..210..383A}
{Aschwanden}, M.~J., {Brown}, J.~C., \& {Kontar}, E.~P. 2002, \solphys, 210,
  383

\bibitem[{{Battaglia} \& {Benz}(2006)}]{2006A&A...456..751B}
{Battaglia}, M., \& {Benz}, A.~O. 2006, \aap, 456, 751

\bibitem[{{Battaglia} \& {Kontar}(2012)}]{2012ApJ...760..142B}
{Battaglia}, M., \& {Kontar}, E.~P. 2012, \apj, 760, 142

\bibitem[{{Bespalov} {et~al.}(1991){Bespalov}, {Zaitsev}, \&
  {Stepanov}}]{1991ApJ...374..369B}
{Bespalov}, P.~A., {Zaitsev}, V.~V., \& {Stepanov}, A.~V. 1991, \apj, 374, 369

\bibitem[{{Bian} {et~al.}(2012){Bian}, {Emslie}, \&
  {Kontar}}]{2012ApJ...754..103B}
{Bian}, N., {Emslie}, A.~G., \& {Kontar}, E.~P. 2012, \apj, 754, 103

\bibitem[{{Bian} {et~al.}(2011){Bian}, {Kontar}, \&
  {MacKinnon}}]{2011A&A...535A..18B}
{Bian}, N.~H., {Kontar}, E.~P., \& {MacKinnon}, A.~L. 2011, \aap, 535, A18

\bibitem[{{Bieber} {et~al.}(1994){Bieber}, {Matthaeus}, {Smith}, {Wanner},
  {Kallenrode}, \& {Wibberenz}}]{1994ApJ...420..294B}
{Bieber}, J.~W., {Matthaeus}, W.~H., {Smith}, C.~W., {et~al.} 1994, \apj, 420,
  294

\bibitem[{{Brown}(1971)}]{1971SoPh...18..489B}
{Brown}, J.~C. 1971, \solphys, 18, 489

\bibitem[{{Brown} {et~al.}(2003){Brown}, {Emslie}, \&
  {Kontar}}]{2003ApJ...595L.115B}
{Brown}, J.~C., {Emslie}, A.~G., \& {Kontar}, E.~P. 2003, \apjl, 595, L115

\bibitem[{{Dennis} {et~al.}(2011){Dennis}, {Emslie}, \&
  {Hudson}}]{2011SSRv..159....3D}
{Dennis}, B.~R., {Emslie}, A.~G., \& {Hudson}, H.~S. 2011, \ssr, 159, 3

\bibitem[{{Dr{\"o}ge}(2000)}]{2000SSRv...93..121D}
{Dr{\"o}ge}, W. 2000, \ssr, 93, 121

\bibitem[{{Emslie}(1980)}]{1980ApJ...235.1055E}
{Emslie}, A.~G. 1980, \apj, 235, 1055

\bibitem[{{Emslie} {et~al.}(2003){Emslie}, {Kontar}, {Krucker}, \&
  {Lin}}]{2003ApJ...595L.107E}
{Emslie}, A.~G., {Kontar}, E.~P., {Krucker}, S., \& {Lin}, R.~P. 2003, \apjl,
  595, L107

\bibitem[{{Fedorenko}(1983)}]{1983SvA....27..640F}
{Fedorenko}, V.~N. 1983, \sovast, 27, 640

\bibitem[{{Fleishman} {et~al.}(2013){Fleishman}, {Kontar}, {Nita}, \&
  {Gary}}]{2013ApJ...768..190F}
{Fleishman}, G.~D., {Kontar}, E.~P., {Nita}, G.~M., \& {Gary}, D.~E. 2013,
  \apj, 768, 190

\bibitem[{{Galeev} \& {Sudan}(1983)}]{1983hppv.book.....G}
{Galeev}, A.~A., \& {Sudan}, R.~N. 1983, {Handbook of plasma physics. Vol. 1:
  Basic plasma physics I.}

\bibitem[{{Ginsburg} \& {Syrovatskii}(1963)}]{1963ICRC....3..301G}
{Ginsburg}, V.~L., \& {Syrovatskii}, S.~I. 1963, in International Cosmic Ray
  Conference, Vol.~3, International Cosmic Ray Conference, 301

\bibitem[{{Guo} {et~al.}(2012){Guo}, {Emslie}, {Kontar}, {Benvenuto},
  {Massone}, \& {Piana}}]{2012A&A...543A..53G}
{Guo}, J., {Emslie}, A.~G., {Kontar}, E.~P., {et~al.} 2012, \aap, 543, A53

\bibitem[{{Holman} {et~al.}(1982){Holman}, {Kundu}, \&
  {Papadopoulos}}]{1982ApJ...257..354H}
{Holman}, G.~D., {Kundu}, M.~R., \& {Papadopoulos}, K. 1982, \apj, 257, 354

\bibitem[{{Holman} {et~al.}(2011){Holman}, {Aschwanden}, {Aurass}, {Battaglia},
  {Grigis}, {Kontar}, {Liu}, {Saint-Hilaire}, \&
  {Zharkova}}]{2011SSRv..159..107H}
{Holman}, G.~D., {Aschwanden}, M.~J., {Aurass}, H., {et~al.} 2011, \ssr, 159,
  107

\bibitem[{{Huang} \& {Li}(2011)}]{2011ApJ...740...46H}
{Huang}, G., \& {Li}, J. 2011, \apj, 740, 46

\bibitem[{{Jakimiec} {et~al.}(1998){Jakimiec}, {Tomczak}, {Falewicz},
  {Phillips}, \& {Fludra}}]{1998A&A...334.1112J}
{Jakimiec}, J., {Tomczak}, M., {Falewicz}, R., {Phillips}, K.~J.~H., \&
  {Fludra}, A. 1998, \aap, 334, 1112

\bibitem[{{Jokipii}(1966)}]{1966ApJ...146..480J}
{Jokipii}, J.~R. 1966, \apj, 146, 480

\bibitem[{{Jokipii} \& {Meyer}(1968)}]{1968PhRvL..20..752J}
{Jokipii}, J.~R., \& {Meyer}, P. 1968, Physical Review Letters, 20, 752

\bibitem[{{Karney}(1986)}]{1986CoPhR...4..183K}
{Karney}, C. 1986, Computer Physics Reports, 4, 183

\bibitem[{{Kennel} \& {Petschek}(1966)}]{1966JGR....71....1K}
{Kennel}, C.~F., \& {Petschek}, H.~E. 1966, \jgr, 71, 1

\bibitem[{{Kontar} {et~al.}(2011{\natexlab{a}}){Kontar}, {Hannah}, \&
  {Bian}}]{2011ApJ...730L..22K}
{Kontar}, E.~P., {Hannah}, I.~G., \& {Bian}, N.~H. 2011{\natexlab{a}}, \apjl,
  730, L22

\bibitem[{{Kontar} {et~al.}(2010){Kontar}, {Hannah}, {Jeffrey}, \&
  {Battaglia}}]{2010ApJ...717..250K}
{Kontar}, E.~P., {Hannah}, I.~G., {Jeffrey}, N.~L.~S., \& {Battaglia}, M. 2010,
  \apj, 717, 250

\bibitem[{{Kontar} {et~al.}(2011{\natexlab{b}}){Kontar}, {Brown}, {Emslie},
  {Hajdas}, {Holman}, {Hurford}, {Ka{\v s}parov{\'a}}, {Mallik}, {Massone},
  {McConnell}, {Piana}, {Prato}, {Schmahl}, \&
  {Suarez-Garcia}}]{2011SSRv..159..301K}
{Kontar}, E.~P., {Brown}, J.~C., {Emslie}, A.~G., {et~al.} 2011{\natexlab{b}},
  \ssr, 159, 301

\bibitem[{{Krucker} {et~al.}(2007){Krucker}, {Kontar}, {Christe}, \&
  {Lin}}]{2007ApJ...663L.109K}
{Krucker}, S., {Kontar}, E.~P., {Christe}, S., \& {Lin}, R.~P. 2007, \apjl,
  663, L109

\bibitem[{{Krucker} \& {Lin}(2002)}]{2002SoPh..210..229K}
{Krucker}, S., \& {Lin}, R.~P. 2002, \solphys, 210, 229

\bibitem[{{Lee}(1982)}]{1982JGR....87.5063L}
{Lee}, M.~A. 1982, \jgr, 87, 5063

\bibitem[{{Lin}(1985)}]{1985SoPh..100..537L}
{Lin}, R.~P. 1985, \solphys, 100, 537

\bibitem[{{Lin} {et~al.}(2002){Lin}, {Dennis}, {Hurford},
  {et~al.}}]{2002SoPh..210....3L}
{Lin}, R.~P., {Dennis}, B.~R., {Hurford}, G.~J., {et~al.} 2002, \solphys, 210,
  3

\bibitem[{{Palmer}(1982)}]{1982RvGSP..20..335P}
{Palmer}, I.~D. 1982, Reviews of Geophysics and Space Physics, 20, 335

\bibitem[{{Peterson} \& {Winckler}(1958)}]{1958PhRvL...1..205P}
{Peterson}, L., \& {Winckler}, J.~R. 1958, Physical Review Letters, 1, 205

\bibitem[{{Petrosian}(2012)}]{2012SSRv..173..535P}
{Petrosian}, V. 2012, \ssr, 173, 535

\bibitem[{{Piana} {et~al.}(2007){Piana}, {Massone}, {Hurford}, {Prato},
  {Emslie}, {Kontar}, \& {Schwartz}}]{2007ApJ...665..846P}
{Piana}, M., {Massone}, A.~M., {Hurford}, G.~J., {et~al.} 2007, \apj, 665, 846

\bibitem[{{Saint-Hilaire} {et~al.}(2008){Saint-Hilaire}, {Krucker}, \&
  {Lin}}]{2008SoPh..250...53S}
{Saint-Hilaire}, P., {Krucker}, S., \& {Lin}, R.~P. 2008, \solphys, 250, 53

\bibitem[{{Schlickeiser}(1989)}]{1989ApJ...336..243S}
{Schlickeiser}, R. 1989, \apj, 336, 243

\bibitem[{{Sim{\~o}es} \& {Kontar}(2013)}]{2013A&A...551A.135S}
{Sim{\~o}es}, P.~J.~A., \& {Kontar}, E.~P. 2013, \aap, 551, A135

\bibitem[{{Skilling}(1975)}]{1975MNRAS.172..557S}
{Skilling}, J. 1975, \mnras, 172, 557

\bibitem[{{Stepanov} \& {Tsap}(2002)}]{2002SoPh..211..135S}
{Stepanov}, A.~V., \& {Tsap}, Y.~T. 2002, \solphys, 211, 135

\bibitem[{{Stepanov} {et~al.}(2007){Stepanov}, {Yokoyama}, {Shibasaki}, \&
  {Melnikov}}]{2007A&A...465..613S}
{Stepanov}, A.~V., {Yokoyama}, T., {Shibasaki}, K., \& {Melnikov}, V.~F. 2007,
  \aap, 465, 613

\bibitem[{{Sturrock}(1968)}]{1968IAUS...35..471S}
{Sturrock}, P.~A. 1968, in IAU Symposium, Vol.~35, Structure and Development of
  Solar Active Regions, ed. K.~O. {Kiepenheuer}, 471

\bibitem[{{Sweet}(1969)}]{1969ARA&A...7..149S}
{Sweet}, P.~A. 1969, \araa, 7, 149

\bibitem[{{Syrovatskii}(1959)}]{1959SvA.....3...22S}
{Syrovatskii}, S.~I. 1959, \sovast, 3, 22

\bibitem[{{Syrovatskii} \& {Shmeleva}(1972)}]{1972SvA....16..273S}
{Syrovatskii}, S.~I., \& {Shmeleva}, O.~P. 1972, \sovast, 16, 273

\bibitem[{{Tautz} {et~al.}(2008){Tautz}, {Shalchi}, \&
  {Schlickeiser}}]{2008ApJ...685L.165T}
{Tautz}, R.~C., {Shalchi}, A., \& {Schlickeiser}, R. 2008, \apjl, 685, L165

\bibitem[{{Vilmer} {et~al.}(2011){Vilmer}, {MacKinnon}, \&
  {Hurford}}]{2011SSRv..159..167V}
{Vilmer}, N., {MacKinnon}, A.~L., \& {Hurford}, G.~J. 2011, \ssr, 159, 167

\bibitem[{{Xu} {et~al.}(2008){Xu}, {Emslie}, \&
  {Hurford}}]{2008ApJ...673..576X}
{Xu}, Y., {Emslie}, A.~G., \& {Hurford}, G.~J. 2008, \apj, 673, 576

\bibitem[{{Zharkova} \& {Gordovskyy}(2006)}]{2006ApJ...651..553Z}
{Zharkova}, V.~V., \& {Gordovskyy}, M. 2006, \apj, 651, 553

\bibitem[{{Zharkova} {et~al.}(2011){Zharkova}, {Arzner}, {Benz}, {Browning},
  {Dauphin}, {Emslie}, {Fletcher}, {Kontar}, {Mann}, {Onofri}, {Petrosian},
  {Turkmani}, {Vilmer}, \& {Vlahos}}]{2011SSRv..159..357Z}
{Zharkova}, V.~V., {Arzner}, K., {Benz}, A.~O., {et~al.} 2011, \ssr, 159, 357

\end{thebibliography}

\appendix
\section{Mean free path for combined collisional and non-collisional scattering}\label{appendix1}

To obtain the form of $D_{\mu\mu}^{(T)}$, consider the equation of motion for electrons
in the magnetostatic approximation :
\begin{equation}\label{eq:motion}
\dot{\mathbf{p}}=\frac{e}{c}\left[\, \mathbf{v}\times (B_{0} \, \mathbf{z} +\delta \mathbf{B}) \, \right] \,\,\, ,
\end{equation}
where the magnetic field ${\mathbf B}$ comprises a background field $B_{0} \, \hat{\mathbf{z}}$ and a fluctuating perpendicular
part $\delta \mathbf{B}$.

The position of the electrons along the loop is determined by the three coordinates $(z,\mu,\phi)$, with $\phi$ being the gyrophase, and $z$ is the
coordinate along the field line and $\mu$ is the cosine of pitch angle; from Equation~(\ref{eq:motion}), these coordinates evolve according to
\begin{equation}
\frac{dz}{dt} = \mu \, v \,\,\, ;
\end{equation}

\begin{equation}
\frac{d\phi}{dt}= \Omega_{ce} \left [ 1 -\frac{\mu}{\sqrt{1-\mu^2}} \, \left ( \cos{\phi} \, {\delta B_x (z) \over B_0} + \sin{\phi} \, {\delta B_y (z) \over B_0} \right ) \right ] \,\,\, ;
\end{equation}
and

\begin{equation}
\frac{d\mu}{dt}= \sqrt{1-\mu^2} \, \, \Omega_{ce} \, \left ( \cos{\phi} \, {\delta B_x (z) \over B_0} - \sin{\phi} \, {\delta B_y (z) \over B_0} \right ) \,\,\, ,
\end{equation}
where $\Omega_{ce}= {e B_0}/{m_e c}$ is the electron gyrofrequency. The pitch-angle diffusion coefficient is defined as

\begin{equation}\label{dmumudef}
D_{\mu \mu}^{(T)}= \int \limits_{0}^{\infty} dt \, \langle \, \dot{\mu} (0) \dot{\mu} (t) \, \rangle \,\,\, .
\end{equation}
In the quasilinear approximation, the Lagrangian correlation function $C_{L}(t)=\langle \, \dot{\mu}(0)\dot{\mu}(t) \, \rangle$ is computed from the unperturbed orbits of the particles. This yields

\begin{eqnarray}
\nonumber C_{L}(t) & = & \langle \, \dot{\mu} (0) \dot{\mu} (t) \, \rangle \\
\nonumber & = & \int dz \, d\phi \, \langle \, \dot{\mu} (0,0) \, \dot{\mu} (z,\phi) \, \delta (z -z(t)) \, \delta (\phi -\phi(t)) \, \rangle \\
& = & \frac{2 \, \Omega^2_{ce}}{B^2_0} \, \int dz \, d\phi \, (1-\mu^2) \, \langle \, \cos{\phi} \,
{\delta B} (0) \, {\delta B} (z) \, \delta (z -z(t)) \, \delta (\phi -\phi(t)) \, \rangle \,\,\, .
\end{eqnarray}
Substituting the unperturbed ($\delta B=0$) values of $z$ and $\phi$, i.e., $z(t)= \mu \, v \, t, \phi(t)= \Omega_{ce} \, t$,
we obtain

\begin{equation}\label{clt}
\nonumber C_{L} (t) = \frac{2 \, \Omega^2_{ce}}{B^2_0} \, (1-\mu^2) \, \langle \, \delta B (0) \, \delta B (\mu v t) \, \rangle \, \cos{\Omega_{ce}t}  \,\,\, .
\end{equation}
Defining the Eulerian correlation function of the magnetic perturbations as

\begin{equation}
C(z)=  \langle \delta B (0) \, \delta B (z) \rangle \,\,\, ,
\end{equation}
we see from Equations~(\ref{dmumudef}) and~(\ref{clt}) that the diffusion coefficient in pitch-angle space can be written as
\begin{equation}
D_{\mu \mu}^{(T)}=  \frac{2 \Omega^2_{ce}}{B^2_0} \int \limits_{0}^{\infty} dt \, (1-\mu^2) \, \cos{(\Omega_{ce}t)} \, C(\mu v t) \,\,\, .
\end{equation}

The standard quasilinear result for slab turbulence \citep{1966ApJ...146..480J,1966JGR....71....1K,1975MNRAS.172..557S} for an arbitrary spectrum of turbulence $W(k_\parallel)$ has the form \citep{1982JGR....87.5063L}

\begin{equation}\label{eq:D_QLT}
D_{\mu\mu}^{(T)}=\frac{\pi}{2} \, (1-\mu^2) \, \Omega _{ce} \left.\frac{k_\parallel W(k_\parallel)}{B_0^2}\right|_{k_\parallel=\Omega _{ce}/v|\mu| },
\end{equation}
where $W(k_\parallel)$, with $k_\parallel$ being the parallel wavenumber, is the spectral energy density of magnetic
fluctuations, normalized so that the total energy density of fluctuations is $\int \limits _{-\infty}^{\infty}W(k_\parallel) \, dk_\parallel=(\delta B)^2$.

The pitch-angle scattering coefficient $D_{\mu\mu}^{(T)}$ is thus dependent on the spectrum of magnetic fluctuations $W(k_\parallel)$, or, equivalently, on the form the correlation function $C(z)$.  In interplanetary space the spectrum of magnetic fluctuations is normally approximated as a power-law. For solar wind conditions,
the quasilinear result given by (\ref{eq:D_QLT}) tends to over-estimate the scattering of particles for the parameters of turbulence in the solar wind and a number of theories has been put forward to improve the expression for $D_{\mu\mu}^{(T)}$ and explain the discrepancies \citep[][]{1982RvGSP..20..335P,1994ApJ...420..294B,2000SSRv...93..121D}.

It must be noted that the spectrum of magnetic fluctuations $W(k_\parallel)$, or the correlation function $C(z)$, is generally unknown in solar flares.  However, as an example, let us assume an exponential correlation function $C(z) \propto \exp \left (- {z}/{\lambda_{B}} \right )$, where $\lambda_{B}$ is the parallel correlation length for magnetic field fluctuations. The corresponding spectrum of magnetic field fluctuations has the Lorentzian form

\begin{equation}
W(k_\parallel) =  \frac{(\delta B)^2}{\pi} \frac{(1/ \lambda_{B})}{(1/ \lambda_{B})^2 + k_\parallel^2} \,\,\, ,
\end{equation}
and the corresponding diffusion coefficient in pitch-angle space becomes

\begin{equation}
D_{\mu \mu}^{(T)}= \frac{1}{2} \, (1-\mu^2) \, \Omega^2_{ce} \, \left ( \frac{\delta B}{B_0} \right )^2 \, \frac{|\mu| v/\lambda_B}{\Omega^2_{ce}+\left(\mu v/\lambda_B \right)^2} \,\,\, .
\end{equation}
In the high-magnetic-field limit $v << \Omega_{ce} \lambda_B$, this can be further simplified to

\begin{equation}\label{dmumu}
D_{\mu \mu}^{(T)}= \frac{|\mu| }{2} \, (1-\mu^2) \, \left ( \frac{\delta B}{B_0} \right )^2  \, \frac{v}{\lambda_{B}} \,\,\, .
\end{equation}

Substitution of Equation (\ref{dmumu}) into Equation~(\ref{eq:lambda_QLT}) of the text yields an infinite mean free path $\lambda$ due to the logarithmic divergence of the integral at the origin; this is a well-known artifact of the approximations employed \citep[e.g.,][]{2008ApJ...685L.165T}. However, $\lambda$ becomes finite when the magnetic fluctuations have non-zero velocity \citep{1983SvA....27..640F,1989ApJ...336..243S},
or when the Lagrangian correlation function $C_{L}(t)=\langle \, \dot{\mu}(0)\dot{\mu}(t) \, \rangle$ is computed from the perturbed orbits of the particles, or when the resonance between particles and magnetic fluctuations is broadened \citep[][]{1982RvGSP..20..335P,1994ApJ...420..294B,2000SSRv...93..121D,2012ApJ...754..103B}.

Here we also note that for binary collisions, $D_{\mu\mu}^{(C)}|_{\mu=0}\neq 0$ and hence a finite $\lambda$ is
obtained when collisions are taken into account.
The  mean free path of a particle undergoing pitch angle scattering due to both binary collisions and magnetic fluctuations is

\begin{equation}\label{eq:A1}
\lambda = \frac{3v}{8} \int _{-1}^{1} \frac{(1-\mu^2)^2}{D_{\mu\mu}^{(T)}+D_{\mu\mu}^{(C)}} \, d\mu \,\,\, .
\end{equation}
Using the pitch angle scattering coefficients (\ref{eq:D_CC}) and~(\ref{dmumu}), one finds

\begin{equation}\label{eq:A2}
\lambda \equiv\frac{3v}{4} \int _{0}^{1} \frac{(1-\mu^2)}{a+b\mu}d\mu \,\,\, ,
\end{equation}
where

$$a=\frac{(1+\overline{Z^2}) K n(z)}{m_e^2 \, v^3}$$
and

$$b=\frac{1}{2} \left ( \frac{\delta B}{B_0} \right )^2  \, \frac{v}{\lambda_{B}} \,\,\, .$$
Performing the integration over $\mu$ in Equation~(\ref{eq:A1}),

\begin{equation}\label{eq:A3}
\lambda =\frac{3v}{4b} \left(\left[1-\frac{a^2}{b^2}\right]\ln \left[\frac{a+b}{a}\right]-\frac{b-2a}{2b}\right) \,\,\, .
\end{equation}
In the case of strong non-collisional scattering $b\gg a$ (i.e., non-collisional scattering operates on shorter scales than does collisions), the expression (\ref{eq:A3}) can be simplified to yield

\begin{equation}\label{eq:A4}
\lambda \simeq\frac{3v}{8} \, \frac{1}{b}\left(2\ln \left[\frac{b}{a}\right]-1\right)=\frac{3 \lambda _B }{4} \, \left ( \frac{B_0}{\delta B} \right )^2
\left(2\ln\left[\frac {m_e^2 \, v^4}{2 \, (1+\overline{Z^2})\, Kn \, \lambda _B} \left ( {\delta B \over B_0} \right )^2 \right] -1 \right).
\end{equation}
In the opposite limit $b\ll a$, $\lambda \simeq {v}/{(2a)}$.

\end{document}